# Quantum computation with two-dimensional graphene quantum dots[*]

Jason Lee(李杰森), Zhi-Bing Li(李志兵), and Dao-Xin Yao (姚道新)[†]

State Key Laboratory of Optoelectronic Materials and Technologies, School of Physics and Engineering, Sun Yat-sen University, Guangzhou 510275, China

## Abstract

We study an array of graphene nano sheets that form a two-dimensional $S = 1/2$ Kagome spin lattice used for quantum computation. The edge states of the graphene nano sheets are used to form quantum dots to confine electrons and perform the computation. We propose two schemes of bang-bang control to combat decoherence and realize gate operations on this array of quantum dots. It is shown that both schemes contain a great amount of information for quantum computation. The corresponding gate operations are also proposed.

## 1. Introduction

There has been increasing interest in grapheme since its discovery. [1–3] It has shown excellent electronic[4,5] and mechanical [6,7] properties and is also a promising candidate for biosensors.[8,9] Before this amazing discovery, Wallace had studied the band structure of graphite and found a linear dispersion around the Dirac point in the Brillouin zone. [10] Much research has been done on this linear dispersion and in particular on the transport properties of graphene. [11−13] Nakada and Fujita studied the edge state and the nano size effect of graphene, and found that the charge can be localized in the zigzag edge[14] to form quantum dots (QDs).

The graphene-based QD has been reported to be a promising candidate for nanoelectronic devices. [15−17] References [18] and [19] have exploited the edge state for quantum information processing (QIP) especially. The weak spin–orbit coupling and the hyperfine interaction in graphene make it desirable for the coherent control of the spin degree of freedom for spin-based quantum computation. [19] In their study, a one-dimensional antiferromagnetic (AFM) spin chain was constructed by the edge state of a graphene nanoribbon to perform quantum computation.

---

[*] Project supported by the National Natural Science Foundation of China (Grant No. 11074310), the National Basic Research Program of China (Grant No. 2007CB935501), and Fundamental Research Funds for the Central Universities of China.
[†] Corresponding author. E-mail: yaodaoxin@gmail.com

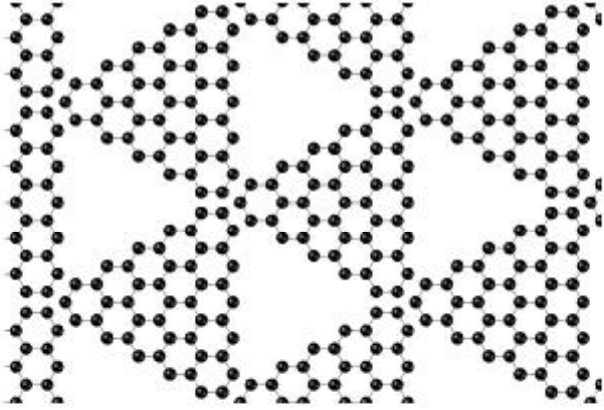

Fig. 1. Two-dimensional array of triangular graphene sheets used for quantum computation.

Here, we consider the grapheme-based system shown in Fig. 1. It is composed of an array of triangular graphene sheets. Each sheet contains three zigzag edges to form QDs. Even though grapheme QDs are affected by ripples in graphene, [20] we do not consider the effect of ripples in our study. All these QDs interact with each other through Heisenberg coupling and build a two-dimensional spin lattice. This structure can be used as a new scheme for QIP. This paper is organized as followings. In Section 2, QDs in graphene are introduced to form a two-dimensional Heisenberg model. In Section 3, two schemes of quantum computation are proposed. Conclusions are given in Section 4.

## 2. Model of quantum dots on graphene

It is well known that the edge state of grapheme exists only in the zigzag edge of a graphene nanoribbon. The combination of zigzag and armchair edges can confine electrons around the zigzag section, as proposed by Guo et al.[4] Here, we combine these two edges in a different way to form confined electrons (QDs), as shown in Fig. 1. Since our system is consisted of an array of triangular graphene sheets with zigzag edges, we expect the QDs to be localized in the edges of the sheets.

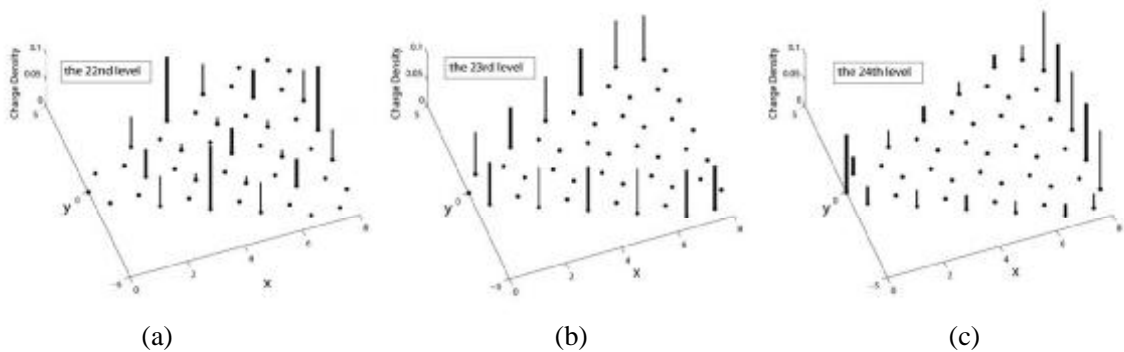

(a)          (b)          (c)

Fig. 2. Distributions of electrons for energy levels (a) just below the Fermi level, (b) and (c) at the Fermi level. Thick and thin bars correspond to different signs of wave function.

The tight binding approximation (TBA)[21,22] is used to describe electrons in graphene. The hopping integrals are chosen to be 2.8 eV, 0.12 eV, and 0.068 eV for the nearest, the second nearest, and the third nearest neighbours, respectively.[18] Thus, the Hamiltonian reads

$$\mathcal{H} = \sum_{L_1, L_2} t(L_1 - L_2) C_{L_1}^\dagger C_{L_2} \quad (1)$$

where $t(L_1 - L_2)$ represents the hopping integral between carbon atoms at sites $L_1$ and $L_2$, and $C_{L_2}$ is the electron destruction operator at site L2. Some studies (e.g. Ref. [18]) have shown that electrons can be confined at room temperature to form QDs if ground state energy $E_0$ and energy gap $E_1$ reach 0.1 eV. These conditions are satisfied for the system under consideration, as $E_0$ = 0.1152 eV and $E_1$ = 0.0996 eV. For our system, each sheet contains 46 carbon atoms, so the diagonalization of the TBA Hamiltonian yields 46 energy levels. The 23-rd and the 24-th ones are the Fermi level, which are degenerate. Figure 2 shows the charge distribution for the three energy levels around the Fermi energy. Figure 2(a) shows the charge distribution of the 21st energy level. Figures 2(b) and 2(c) show the charge distributions of the 23rd and the 24$^{th}$ energy levels, respectively. After some linear combinations, we obtain the charge distributions of the QDs shown in Fig. 3. Figures 2 and 3 show clearly that these three QDs exist on all the three sides of the triangular graphene sheets. The wave functions can be combined to obtain energy levels that are similar to that of cycle propane.

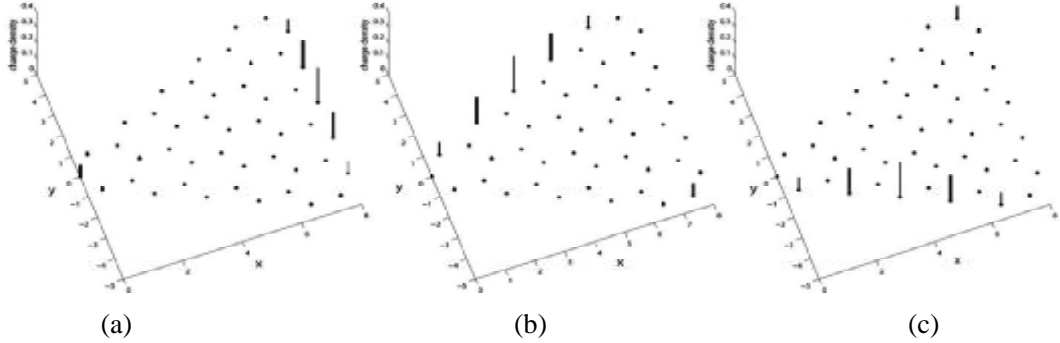

(a)            (b)            (c)

Fig. 3. Panels (a), (b), and (c) exhibit the distribution of electron density for the three QDs. Thick and thin bars correspond to different signs of wave function

From the distribution of the QDs, we can calculate the spin interaction between these QDs by using the Hubbard approximation. In the following calculation, the on-site Coulomb interaction in a carbon atom is taken to be 3.5 eV, [23] and the overlap integral between two neighboring carbon atoms is chosen to be 0.129 eV. [24] Since the overlap integral decreases exponentially with the distance, we assume that the overlap integral between two carbon atoms has the form of **$0.129^L$**, where $L$ is the distance divided by 1.42Å. From these, we can obtain hopping TQD between two QDs via the formula

$$T_{QD} = E_0 \int \psi_1^*(r) \psi_2(r) \mathrm{d}^3 r \quad (2)$$

where $\psi_1(r)$ and $\psi_2(r)$ are the wave functions of the two QDs, respectively, and E0 is the ground state energy. From the TBA calculation, we obtain

$$\psi_i(r) = \sum_L a_i(L)\phi(r-L) \tag{3}$$

where $\phi(r)$ is the p-orbital wave function of the carbon atom, and ai(L) are the coefficients in the linear combination of atomic orbitals (LACO). The effective Hubbard $U$ of the QD can be written as

$$U_{QD} = \iint \psi^*(r_1)\psi(r_1)\frac{1}{|r_1-r_2|}\psi^*(r_2)\psi(r_2)d^3r_1 d^3r_2 \tag{4}$$

The condition $U_{QD} \gg T_{QD}$ is satisfied in our case, justifying the use of the Hubbard approximation. Thus, the spin exchange coupling J can be calculated by

$$J = \frac{1}{2}\sqrt{T_{QD}^2 + U_{QD}^2} - \frac{1}{2}U_{QD} \approx \frac{4T_{QD}^2}{U_{QD}} \tag{5}$$

The spin exchange interaction between two QDs forms a Heisenberg-type Hamiltonian, $\mathcal{H}_{12} = J_{12}S_1 \cdot S_2$. Here $J$ is always positive according to Eq. (5), meaning that the interaction is antiferromagnetic (AFM). Therefore, we obtain an asymmetric Kagome spin lattice, which has spin coupling interactions of $J_1$ =6.27 μeV and $J_2$ = 2.68 μeV, as shown in Fig. 4. In the case of $J_1 = J_2$, this asymmetric Kagome lattice is reduced to the uniform Kagome lattice.

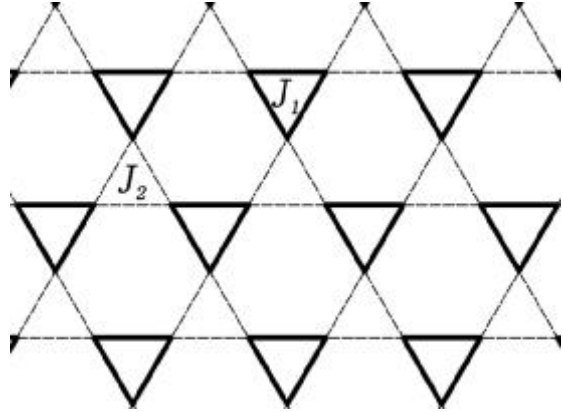

Fig. 4. Scheme diagram of the triangular spin system. The thick solid lines represent the interaction between the QDs in the same triangular graphene sheet, and the thin dashed lines represent the interactions between the QDs in two neighboring sheets. Here $J_1$= 6.27 μeV and $J_2$= 2.68 μeV.

## 3. Quantum computation based on quantum dots

In QIP, decoherence is a major issue. If improperly handled, information cannot be stored in the device, which renders QIP impossible. Based on its natural abundance, about 1% of the C atoms in grapheme are 13C atoms.[25] From the data, we can obtain the decoherence time, which is in the order of 10μs[25] and is several orders of magnitude larger than the operation time in our system ($\pi\hbar/4J$=0.19 ns[18]). This guarantees our quantum computation time. In Ref. [26]. the following Hamiltonian is

considered:
$$\mathcal{H} = \mathcal{H}_S \otimes I_B + I_S \otimes \mathcal{H}_B + \mathcal{H}_{SB} \tag{6}$$
where S and B refer to system and bath, respectively, and $I_S$ and $I_B$ are the corresponding unit operators. The $\mathcal{H}_{SB}$ is the interaction between system and bath, and can be expressed as
$$\mathcal{H}_{SB} = \sigma_x \otimes b_x + \sigma_y \otimes b_y + \sigma_z \otimes b_z \tag{7}$$
where $b_i$ ($i = x, y, z$) refers to the bath Hamiltonian, and $\sigma_i$ ($i = x, y, z$) are the spin operators. A sequence of fast and strong pulses can be used to form a unitary group operation $\{g\}$, which modifies the original Hamiltonian H as[26]
$$\mathcal{H} \mapsto \mathcal{H}_{eff} = \frac{1}{|G|} \sum_{g \in G} g^\dagger \mathcal{H} g \tag{8}$$
where $\mathcal{H}_{eff}$ is the effective Hamiltonian. We can apply a similar technique to our asymmetric Kagome lattice.

## 3.1. Scheme one

For the asymmetric Kagome lattice, group operation $\{I, -\sigma_z^1 \otimes \sigma_z^2\}$ is insufficient for QIP. In this scheme, the operation group for each spin becomes $\{I, -i\sigma_x, -i\sigma_y, -i\sigma_z\}$ and is staggered for each particular pulse, as shown in Fig 5(a). The effective operation of $\sigma_x$ for each spin becomes
$$\sigma_x \mapsto \frac{1}{4} \left[ I^\dagger \sigma_x I + (-i\sigma_x)^\dagger \sigma_x (-i\sigma_x) + (-i\sigma_y)^\dagger \sigma_x (-i\sigma_y) + (-i\sigma_z)^\dagger \sigma_x (-i\sigma_z) \right]$$
$$= 0 \tag{9}$$
Thus, term $\sigma_x \otimes b_x$ in $\mathcal{H}_{SB}$ vanishes. Similarly, $\sigma_y \otimes b_y$ and $\sigma_z \otimes b_z$ vanish too, making the system totally decoupled from the environment. We note that the coupling of spin and environment for each spin does not have to be the same, since the images of the Pauli matrices in the bang-bang map vanish.

For the system that does not interact with the environment, we need to address the Heisenberg interactions between the neighboring spins. Since the operations on the neighboring spins are different, the group operation becomes
$$\{I, (-i\sigma_x) \otimes (-i\sigma_y), (-i\sigma_y) \otimes (-i\sigma_z), (-i\sigma_z) \otimes (-i\sigma_x)\}$$
$$= \{I, -\sigma_x \otimes \sigma_y, -\sigma_y \otimes \sigma_z, -\sigma_z \otimes \sigma_x\}$$
then
$$\sigma_x \otimes \sigma_x \mapsto \frac{1}{4} \Big[ I^\dagger \sigma_x I + (-\sigma_x \otimes \sigma_y)^\dagger (\sigma_x \otimes \sigma_x)(-\sigma_x \otimes \sigma_y)$$
$$+ (-\sigma_y \otimes \sigma_z)^\dagger (\sigma_x \otimes \sigma_x)(-\sigma_y \otimes \sigma_z)$$
$$+ (-\sigma_z \otimes \sigma_x)^\dagger (\sigma_x \otimes \sigma_x)(-\sigma_z \otimes \sigma_x) \Big] = 0, \tag{10}$$

which means that the x component of the Heisenberg interaction vanishes under the bang-bang control. We can obtain the same results for $\sigma_y \otimes \sigma_y$ and $\sigma_z \otimes \sigma_z$. Consequently, all the QDs become decoupled. So far, we obtain an array of spins lined up in a triangle, which do not interact with each other under the bang-bang control. Since these spins evolve independently, the encoded states can be chosen as the spin orientations, and the traditional qubit operations are applicable to this system.[27] In this case, each spin acts as one qubit, which means that we have a bit of information per spin. To retrieve the information, we need to pinpoint the locations of the corresponding QDs and measure their spin states. Experimentally, we can use a spin-polarized scanning tunneling microscope (SP-STM) to realize the operations.

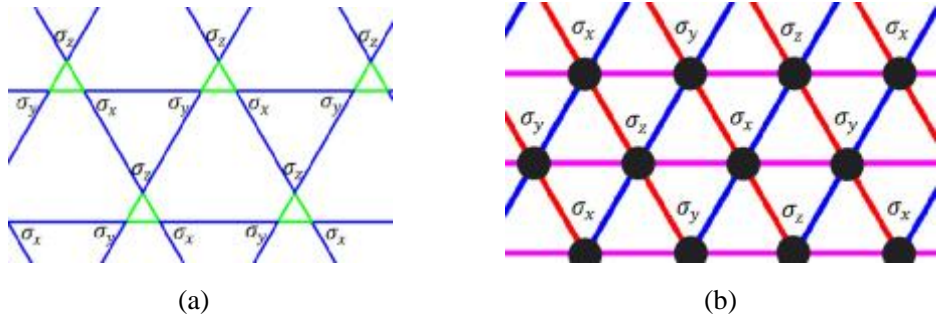

(a)          (b)

Fig. 5. (colour online) One of the operations of the bang-bang control in our model for (a) scheme one and (b) scheme two. Each interception in panel (a) represents one QD, while that in panel (b) represents three QDs in the same graphene sheet. The other two operations can be obtained by rotating indexes $\{x, y, z\}$. The unit operation does not change anything (replacing all $\sigma$'s by $I$)

### 3.2. Scheme two

Unlike scheme one, we propose scheme two to treat the three spins in the same triangular grapheme sheet identically, as shown in Fig. 5(b). Similarly, the system-bath coupling can be eliminated (Eq. (9)), and the interactions between the QDs in different graphene sheets can be eliminated too (Eq. (10)). However, Heisenberg interactions still exist in the same graphene sheets, since

$$\sigma_x \otimes \sigma_x \mapsto \frac{1}{4}\big[I^\dagger \sigma_x I + (-\sigma_x \otimes \sigma_x)^\dagger (\sigma_x \otimes \sigma_x)(-\sigma_x \otimes \sigma_x)$$

$$+ (-\sigma_y \otimes \sigma_y)^\dagger (\sigma_x \otimes \sigma_x)(-\sigma_y \otimes \sigma_y)$$

$$+ (-\sigma_z \otimes \sigma_z)^\dagger (\sigma_x \otimes \sigma_x)(-\sigma_z \otimes \sigma_z)\big] = 0, \quad (11)$$

The same result can be obtained for the y and the z components. Consequently, each graphene sheet behaves independently, and the three QDs interact with each other via Heisenberg coupling (Fig. 6),

$$h = J_1(\boldsymbol{\sigma}_1 \cdot \boldsymbol{\sigma}_2 + \boldsymbol{\sigma}_2 \cdot \boldsymbol{\sigma}_3 + \boldsymbol{\sigma}_3 \cdot \boldsymbol{\sigma}_1) \quad (10)$$

Diagonalization of this Hamiltonian yields 6 degenerate ground states yields

$$\frac{1}{\sqrt{2}}(|\uparrow\downarrow\rangle - |\downarrow\uparrow\rangle) \otimes |\uparrow\rangle$$

$$\frac{1}{\sqrt{2}}(|\uparrow\downarrow\rangle - |\downarrow\uparrow\rangle) \otimes |\downarrow\rangle$$

$$\frac{1}{\sqrt{2}}(|\uparrow_1\downarrow_3\rangle - |\downarrow_1\uparrow_3\rangle) \otimes |\uparrow_2\rangle \qquad (11)$$

$$\frac{1}{\sqrt{2}}(|\uparrow_1\downarrow_3\rangle - |\downarrow_1\uparrow_3\rangle) \otimes |\downarrow_2\rangle$$

$$\frac{1}{\sqrt{2}}|\uparrow\rangle \otimes (|\uparrow\downarrow\rangle - |\downarrow\uparrow\rangle)$$

$$\frac{1}{\sqrt{2}}|\downarrow\rangle \otimes (|\uparrow\downarrow\rangle - |\downarrow\uparrow\rangle)$$

Obviously, these states are the Descartes products of two states, with one being the maximum entangled state between two spins and the other being a third spin that is not entangled. These six states are linearly dependent, and the number of independent states is actually four. If we can make full use of these four states, the information capacity should be $\log_2 4 = 2$ per three QDs, i.e., two qubits (QBs) for three QDs.

Table 1. Codes for quantum computation in system with four degenerate ground states.

| QB1 | QB2 | 0 | 1 |
|---|---|---|---|
| | 0 | $\frac{1}{\sqrt{2}}|\uparrow\rangle \otimes (|\uparrow\downarrow\rangle - |\downarrow\uparrow\rangle)$ | $\frac{1}{\sqrt{2}}(|\uparrow\downarrow\rangle - |\downarrow\uparrow\rangle) \otimes |\uparrow\rangle$ |
| | 1 | $\frac{1}{\sqrt{2}}|\downarrow\rangle \otimes (|\uparrow\downarrow\rangle - |\downarrow\uparrow\rangle)$ | $\frac{1}{\sqrt{2}}(|\uparrow\downarrow\rangle - |\downarrow\uparrow\rangle) \otimes |\downarrow\rangle$ |

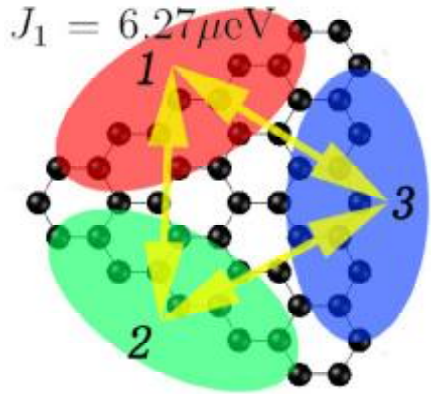

Fig. 6. (colour online) The Heisenberg interactions between three QDs in the same graphene sheet. The coloured ovals represent the three QDs, while the arrowed (yellow) lines denote the interactions between the QDs.

The choice of encoded states for two qubits involving three spins may not be easy if the spins do not act independently as in the scheme one case, and the encoded states are more complicated than that of the free spins. Fortunately, any four of these ground states are independent. We choose the codes shown in Table 1. It is clear that the first qubit is determined by the total spin, whereas the second qubit is determined by the free spin. Then we can choose the operations in Table 2 for gate manipulations. For instance, the X gate on qubit 1 (2) flips the first (second) qubit from 0 to 1, and vice versa; the controlled not gate CNOT1 (CNOT2) flips the second qubit if and only if the first qubit is 0 (1). Figure 7 shows the corresponding operation of CNOT1 as an example of the operations listed in Table 2.

Table 2. Gate operations for quantum computation ($i$ = 0, 1).

| Gate | Location | Operation | Result |
|---|---|---|---|
| Z | QB1 | $Z_1 = \sigma_1^z \sigma_2^z \sigma_3^z$ | $Z_1\|0i\rangle = \|0i\rangle$ <br> $Z_1\|1i\rangle = -\|1i\rangle$ |
| Z | QB2 | $Z_2 = \frac{1}{2}(\sigma_1^z + \sigma_3^z - i\sigma_1^x \sigma_3^y + i\sigma_2^y \sigma_3^x - \sigma_1^z \sigma_2^x \sigma_3^x + \sigma_1^x \sigma_2^x \sigma_3^z)$ | $Z_2\|i0\rangle = \|i0\rangle$ <br> $Z_2\|i1\rangle = -\|i1\rangle$ |
| X | QB1 | $X_1 = -\sigma_1^x \sigma_2^x \sigma_3^x$ | $X_1\|0i\rangle = \|1i\rangle$ <br> $X_1\|1i\rangle = -\|0i\rangle$ |
| X | QB2 | $X_2 = -\frac{1}{2}(1 + \boldsymbol{\sigma}_1 \cdot \boldsymbol{\sigma}_3)$ | $X_2\|i0\rangle = \|i1\rangle$ <br> $X_2\|i1\rangle = \|i0\rangle$ |
| Hadamard | QB1 | $H_1 = \frac{1}{\sqrt{2}}(1 - i\sigma_1^y \sigma_2^y \sigma_3^y)$ | $H_1\|0i\rangle = \frac{1}{\sqrt{2}}(\|0i\rangle + \|1i\rangle)$ <br> $H_1\|0i\rangle = \frac{1}{\sqrt{2}}(-\|0i\rangle + \|1i\rangle)$ |
| Hadamard | QB2 | $H_2 = \frac{1}{2\sqrt{2}}(-1 + \sigma_1^z + \sigma_3^z - \boldsymbol{\sigma}_2 \cdot \boldsymbol{\sigma}_3 - \boldsymbol{\sigma}_1 \cdot \boldsymbol{\sigma}_2 - i\sigma_1^x \sigma_2^y + i\sigma_2^y \sigma_3^x - \sigma_1^z \sigma_2^x \sigma_3^x + \sigma_1^x \sigma_2^x \sigma_3^x)$ | $H_1\|00\rangle = \frac{1}{\sqrt{2}}(\|00\rangle + \|01\rangle)$ <br> $H_1\|01\rangle = \frac{1}{\sqrt{2}}(-\|00\rangle + \|01\rangle)$ <br> $H_1\|0i\rangle = \frac{1}{\sqrt{2}}(-\|10\rangle + \|11\rangle)$ <br> $H_1\|0i\rangle = \frac{1}{\sqrt{2}}(\|10\rangle - \|11\rangle)$ |
| CNOT1 | QB1 and QB2 | $C_1 = -Z_1\left[1 + \frac{1}{4}(\boldsymbol{\sigma}_2 \cdot \boldsymbol{\sigma}_3 - 1)(1 + \sigma_2^z)\right]$ | $C_1\|00\rangle = \|01\rangle$ <br> $C_1\|01\rangle = \|00\rangle$ <br> $C_1\|10\rangle = \|10\rangle$ <br> $C_1\|11\rangle = \|11\rangle$ |
| CNOT2 | QB1 and QB2 | $C_2 = -Z_1\left[1 + \frac{1}{4}(\boldsymbol{\sigma}_2 \cdot \boldsymbol{\sigma}_3 - 1)(1 - \sigma_2^z)\right]$ | $C_2\|00\rangle = \|00\rangle$ <br> $C_2\|01\rangle = \|01\rangle$ <br> $C_2\|10\rangle = \|11\rangle$ <br> $C_2\|11\rangle = \|10\rangle$ |

We have to admit that the Hadamard gate on the second qubit has some problems, since, as mentioned above, the encoded states are more complicated than the free spins. It is not exactly the Hadamard gate that one would normally expect, and further work still needs to be done. We notice that there are some interesting operations, such as operation $\frac{1}{2}(1 + \sigma_1 \cdot \sigma_2)$, which have not been utilized. The above operation acts like a Hadamard gate when the second qubit is 0 ($|i0\rangle \to |i0\rangle + |i1\rangle, i = 0,1$), while it behaves as a Z gate when the second qubit is 1 ($|i1\rangle \to -2|i1\rangle$). Similarly, $\frac{1}{2}(1 + \sigma_2 \cdot \sigma_3)|i1\rangle = |i0\rangle + |i1\rangle$, and $\frac{1}{2}(1 + \sigma_2 \cdot \sigma_3)|i0\rangle = -2|i0\rangle, i = 0,1$. These operations may be useful in quantum computation. Here we have information of 2 ($\log_2 4 = 2$) instead of 2 ($\log_2 2 = 2$), which is less than that in scheme one. Since the three QDs in the same grapheme sheets are treated identically, it may be easier to realize experimentally than scheme one. In scheme two, two qubits are stored in the entangled states of the corresponding three QDs. This means that we need to single out the graphene sheet in which the three QDs are located, and then measure the spin distribution of the graphene sheet. More specifically, if the total spin is $\frac{1}{2}$ ($-\frac{1}{2}$), then the first qubit is 0 (1); if the spin is polarized in the first (third) QD, then the second qubit is 0 (1). A spin-polarized STM can be used to realize the operation. Because of the periodicity of the lattice, we can do the operation in the Fourier transformed space, which gives advantages to our schemes.

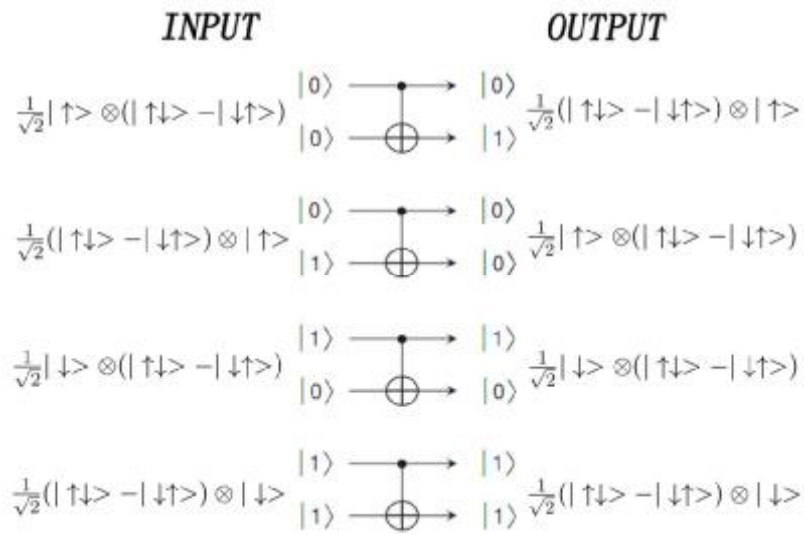

Fig. 7. Operations of controlled X gate (CNOT) on qubit one. The upper qubit acts as a controller determining whether the lower qubit should be flipped or not.

## 4. Conclusion

In conclusion, we have suggested two possible schemes to realize quantum computation with an array of triangular graphene sheets. The array is constructed with equilateral triangles of graphene sheets with three zigzag edges to make use of the edge states of the zigzag honeycomb lattice. The graphene sheets give rise to an array of graphene quantum dots, which interact with each another via Heisenberg coupling. Our calculations predict the couplings between the QDs to be $J_1$ = 6.27 μeV and $J_2$ = 2.68 μeV for spins staying in the same graphene sheet and in neighboring graphene sheets, respectively. These quantum dots form an asymmetric Kagome Heisenberg model with S = 1/2, which is closely related to the Heisenberg model in the Kagome lattice. For the asymmetric Kagome spin system, two schemes of the bang-bang control solutions for the quantum computation are put forward, both of which have a sequence of pulse operations imposed on the system to eliminate the coupling with the environment as well as to modify the Hamiltonian to make quantum computation possible. For scheme one, the encoded states can be chosen as the spin states, and the gate operations can be realized by the spin flippings. In this scheme, each spin contains a unit amount of information. In scheme two, each spin interacts with two other spins in the same graphene sheet through Heisenberg coupling while having no interaction with other spins in different sheets. In scheme two, the corresponding Hamiltonian is the direct product of the Hamiltonian of the three interacting spins, which yields four degenerate ground states. The choice of encoded states based on the four degenerate ground states and the corresponding gate operations are proposed too. Scheme two gives 2/3 unit of information per qubit. While scheme one handles more information, scheme two may be more easier to realize experimentally. In summary, our schemes with two-dimensional graphene quantum dots contain a larger amount of information than that in the one-dimensional case, which provides a promising solution for graphene-based quantum computation.

## Acknowledgment

We thank Jing-Bo Xu and Tai-Kai Ng for helpful discussions.